\documentclass[twocolumn,prl,aps,floatfix,epsf,psfig,superscriptaddress,amssymb]{revtex4}
\usepackage{graphicx}
\usepackage{amsmath}
\usepackage[colorlinks=true, citecolor=blue, urlcolor=blue, linkcolor=blue,bookmarks=false,hypertexnames=true]{hyperref} 
\usepackage{doi}

\newcommand{\ket}[1]{\left| #1 \right\rangle}

\begin{document}
\graphicspath{}

\title{Spin-valley qubits in gated quantum dots in a single layer of transition metal dichalcogenides}

\author{Abdulmenaf Alt{\i}nta\c{s}}
\affiliation{Department of Physics, University of Ottawa, Ottawa, Ontario, Canada K1N 6N5}

\author{Maciej Bieniek}
\affiliation{Department of Physics, University of Ottawa, Ottawa, Ontario, Canada K1N 6N5}
\affiliation{Department of Theoretical Physics, Wroc\l aw University of Science and Technology, Wybrze\.ze Wyspia\'nskiego 27, 50-370 Wroc\l aw, Poland}

\author{Amintor Dusko}
\affiliation{Department of Physics, University of Ottawa, Ottawa, Ontario, Canada K1N 6N5}

\author{Marek Korkusi\'nski}
\affiliation{Security and Disruptive Technologies, Emerging Technologies Division, NRC, Ottawa}

\author{Jaros{\l}aw Paw{\l}owski}
\affiliation{Department of Theoretical Physics, Wroc\l aw University of Science and Technology, Wybrze\.ze Wyspia\'nskiego 27, 50-370 Wroc\l aw, Poland}

\author{Pawe\l \ Hawrylak}
\affiliation{Department of Physics, University of Ottawa, Ottawa, Ontario, Canada K1N 6N5}

\date{\today}

\begin{abstract}
We develop a microscopic and atomistic theory of electron spin based-qubits in gated quantum dots in a single layer of transition metal dichalcogenides. The qubits are identified with two degenerate locked spin and valley states in a gated quantum dot.  The two qubit states are accurately described using a multi-million atom tight-binding model solved in wavevector space. The spin-valley locking and strong spin-orbit coupling result in two degenerate states, one of the qubit states  being spin down located at the $+K$ valley of the Brillouin zone, and the other state  located at the $-K$ valley with spin up. We describe the qubit operations necessary to rotate the spin-valley qubit as  a combination of the applied  vertical electric field, enabling spin-orbit coupling in a single valley, with a lateral strongly localized valley-mixing gate.
\end{abstract}

\pacs{}
\maketitle

\setcounter{secnumdepth}{3} 
\section{Introduction}
There is currently interest in developing quantum circuits based on electron spin qubits \cite{Brum_Hawrylak_1997, Loss_DiVincenzo_1998, Tarucha_Kouwenhoven_1996, Ciorga_Wasilewski_2000, Meier_Loss_2003, Elzerman_Kouwenhoven_2004, Hanson_Vandersypen_2007, Kyriakidis_Hawrylak_2002, klinovaja2013spintronics} in gated quantum dots in gallium arsenide and silicon \cite{Hsieh_Hawrylak_2012, Nowack_Vandersypen_2011, Botzem_Bluhm_2018, West_Dzurak_2019, Ito_Tarucha_2018}.  In these structures, electrons are localised in a volume containing millions of atoms, hence the nuclear spins and atomic vibrations contribute to the decoherence of electron spins. Recent realization of semiconductor layers with atomic thickness \cite{CastroNeto_Geim_2009,Kadantsev_Hawrylak_2012, Geim_Grigorieva_2013, Ihn_Ensslin_2010, Mak_Heinz_2010, Manzeli_Kis_2017, Guclu_Hawrylak_2014, Scrace_Hawrylak_2015,Tu_Kardynal_2019,Peng_Cheng_2019,Klein_Holleitner_2019,Jadczak_Hawrylak_2019} opens the possibility of confining single electrons to a few-atom thick layers, potentially significantly increasing the operating temperature and the coherence of electron spin qubits. 

Recently, quantum dots (QDs) in transition metal dichalcogenides (TMDCs), graphene, and bilayer graphene have been realized 
\cite{Guttinger_Ensslin_2010, Guclu_Hawrylak_2014, McGuire_2016, Wang_Ruffieux_2017, Wang_Kim_2018, Pisoni_Ensslin_2018}
 by creating electrostatic confinement with lateral metal electrodes \cite{Volk_Stampfer_2011, Allen_Yacoby_2012, Eich_Ensslin_2018, Pisoni_Ensslin_2018, Wang_Kim_2018, Kurzmann_Ihn_2019}. Several groups reported the creation of finite-size electron droplets using metallic gates \cite{Wang_Kim_2018, Pisoni_Ensslin_2018, BrotonsGisbert_Gerardot_2019}. Gated quantum dots combined with large trion binding energies allowed for electrical probing of excitons in TMDC QDs \cite{Lu_Srivastava_2019, Pisoni_Ensslin_2018, Wang_Kim_2018, BrotonsGisbert_Gerardot_2019, Chakraborty_Vamivakas_2018}. For example, a local tunable confinement potential has been realized by Kim and co-workers\cite{Wang_Kim_2018}, and gate tuning of QD molecules have been shown by Guo and co-workers \cite{Zhang_Guo_2017}. There has also been significant progress in theoretical understanding of TMDC QDs. Stability and electronic properties of small QDs with various composition, orientation, and edge type have been studied within DFT (Density Functional Theory) \cite{Lauritsen_Besenbacher_2003,Pei_Song_2015, Javaid_Greentree_2017,  Lauritsen_Besenbacher_2007, McBride_Head_2009, Li_Galli_2007}. In particular, Galli and co-workers \cite{Li_Galli_2007} studied the electronic properties of  triangular MoS$_\textrm{2}$ quantum dots as a function of the number of layers and predicted a transition to a direct gap semiconductor in a single layer. 

Nevertheless, ab-initio approaches are limited to small structures, and to describe quantum dots with lateral sizes up to tens of nanometers, one can make use of hybrid DFT based tight-binding models\cite{Liu_Xiao_2013, Rostami_Asgari_2013, Cappelluti_Guinea_2013, Zahid_Guo_2013, Fang_Kaxiras_2015, Ho_Chen_2014, Liu_Yao_2015, Ridolfi_Lewenkopf_2015, Shanavas_Satpathy_2015, SilvaGuillen_Roldan_2016, Pearce_Burkard_2016, Dias_Fu_2018}. Using a 3-band tight-binding model limited to metal orbitals, Peeters, and co-workers analyzed the effect of quantum dot shape and external magnetic field on the single-particle energy spectrum  \cite{Pavlovic_Peeters_2015, Chen_Peeters_2018}.  Using an atomistic tight-binding approach, spin-valley qubits have been described in small quantum dots by Bednarek and co-workers \cite{Pawlowski_Bednarek_2018, pawlowski2019spin}, Szafran and co-workers \cite{Zebrowski_Szafran_2013, Szafran_Kolasinska_2018, Szafran_Zebrowski_2018} and Guinea and co-workers \cite{Chirolli_Roldan_2019}. Using such an approach, two valley-qubit operations have also been recently proposed by some of us \cite{Pawlowski_Wozniak_2021}. In order to understand the size dependence of the electronic states in quantum dots for realistic sizes involving millions of atoms, $k\cdot p$ and effective massive Dirac fermion models were also applied \cite{Kormanyos_Burkard_2014, Liu_Yao_2014, Brooks_Burkard_2017, Szechenyi_Palyi_2018, Qu_Azevedo_2017, Dias_Qu_2016}.

In our previous work, an ab-initio based tight-binding model combining metal and chalcogen orbitals, applicable to multi-million atom quantum dots in TMDCs, has been developed \cite{Bieniek_Hawrylak_2018}. We note that in a tight binding model the correct level degeneracies occur, but their direct identification with valleys is difficult. By working in reciprocal space, the valleys were explicitly taken into account. The effect of valley, spin, and band nesting on the electronic properties of gated quantum dots in a single layer of transition metal dichalcogenides was described \cite{Bieniek_Hawrylak_2020}, along with valley- and spin-polarized broken-symmetry many-body states discussed in Ref. \onlinecite{Szulakowska_Hawrylak_2020}. It was shown that the lowest electronic state confined in a quantum dot is a doublet of spin and valley locked states. Hence, such a doublet could serve as a qubit. In order to realize a spin-valley qubit, a way to control spin and valley properties of electrons in these QDs is needed. Several means of manipulating the valley index in quantum dots have been already studied: strain \cite{Chirolli_Roldan_2019}, magnetic field \cite{Kormanyos_Burkard_2014, Brooks_Burkard_2017, Szechenyi_Palyi_2018}, and coupling to impurity \cite{Szechenyi_Palyi_2018}. Valley mixing by the confining potential has also been analyzed by Yao and co-workers \cite{Liu_Yao_2014} and the magnetic control of the spin-valley coupled states in TMDC QDs has been shown by Qu and co-workers \cite{Qu_Azevedo_2017,Dias_Qu_2016}. 

In this paper, building on our previous work, we expand our microscopic theory of electron spin-valley qubits and provide a prescription of how to manipulate the two qubit states. The microscopic tight-binding model developed here is suitable for accurate description of multi-million atom nanostructures compatible with existing experiments. 
The two degenerate qubit states, belonging to the two non-equivalent valleys, each with the opposite-spin, are built out of conduction band states of even parity with respect to the metal plane.  The rotation of the qubit, the logical $\sigma_x$ operation, requires simultaneous transition between opposite spin states in each valley and between the two nonequivalent valleys. The understanding of the orbital composition of conduction band states as a linear combination of even parity metal orbitals and even parity sulfur dimer orbitals allows us to show that the qubit rotation  is accomplished by applying both a time dependent vertical electric field and a time dependent highly localized lateral potential. The electric field couples primarily to the two sulfur layers, and activates odd conduction bands, which enables in turn spin flips on metal atoms due to the spin-orbit interaction. The admixture of an opposite-spin orbital and application of a lateral local potential enables transition to the opposite valley and spin qubit state. This process is illustrated in Figure 1. Figure 1 shows a cross section of a schematic device consisting of a single TMDC layer, with metallic gates (shown in yellow ) producing a lateral potential confining a single electron to a quantum dot in a single TMDC layer , illustrated with a thick arrow below. In addition, a metallic vertical gate, implemented here with two graphene layers, generates an on demand vertical electric field. The local gate, implemented here with an STM (Scanning Tunneling Microscope) tip, generates an on demand valley mixing potential. 
The suggested set up shown in Figure 1 is compatible with experimental designs and implementation of gated quantum dot in a single layer of WSe2 \citep{boddison2021gate}. 
We will show that turning these two gates on for a finite time rotates spin valley qubit from logical qubit 0 to logical qubit 1.   

The paper is organized as follows. In  section 2, we describe logical quantum bits encoded in two lowest degenerate states of an electron confined in a lateral gated quantum dot in TMDC. In section 3, we describe the effect of two external gates allowing for flipping of the spin and flipping of the valley, necessary for logical qubit quantum operations. In section 4, we summarise our results.

\section{Logical quantum bits encoded in electronic states of an electron confined in a lateral gated quantum dot}

\begin{figure}[ht]
\centering
\includegraphics[width=0.5\textwidth]{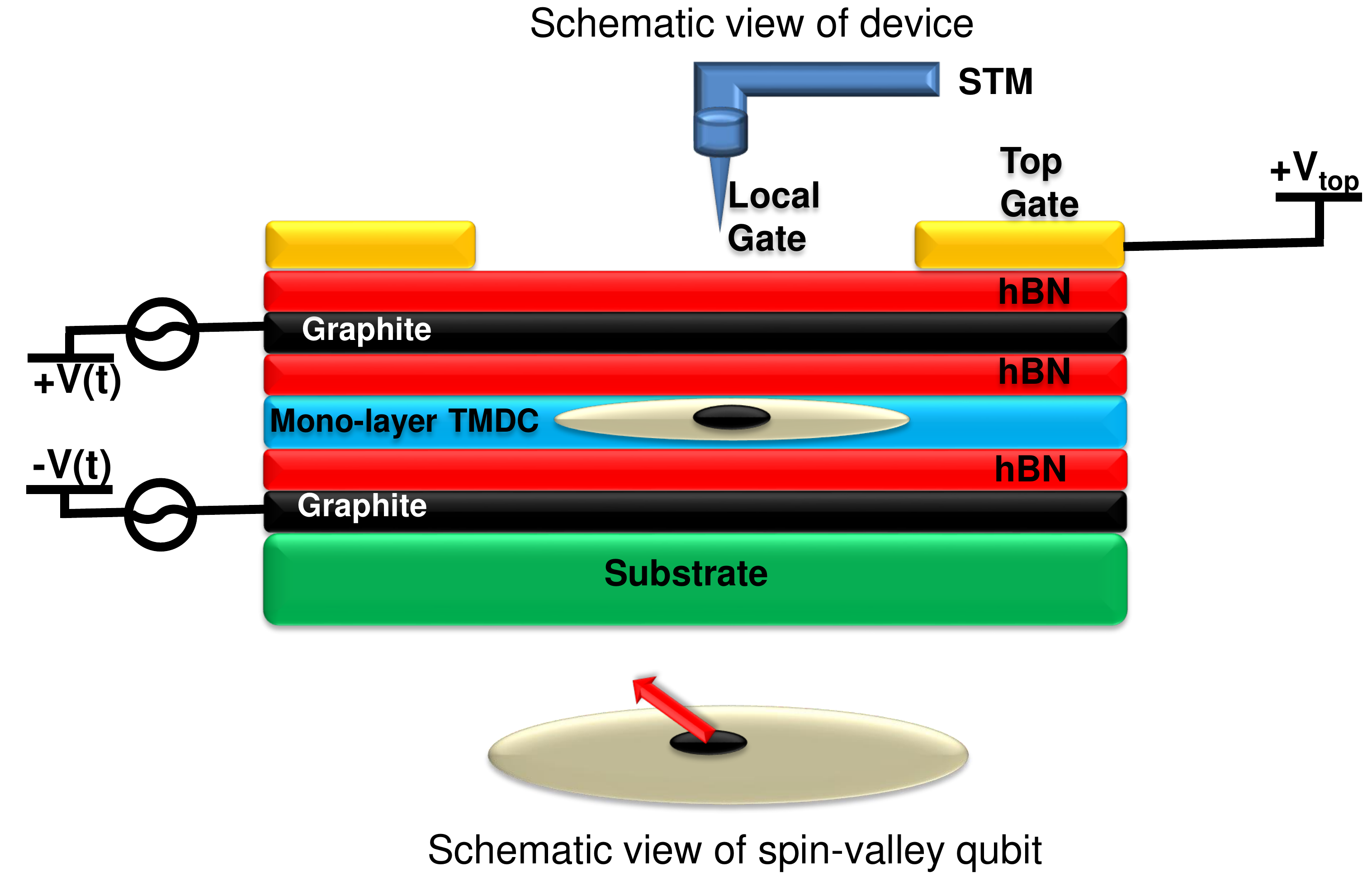}
\caption{(Color online) Top: A schematic view of device. Quantum dot in monolayer TMDC  is induced by top  gate (gold), and highly localized potential necessary for valley mixing is controlled here by scanning  tunneling microscope tip. Additional vertical electric field is induced  by potentials applied to two graphite  layers. Bottom: Schematic view of spin-valley qubit (red arrow).} 
\label{fig1}
\end{figure}

Here we identify and analyse the logical quantum bits encoded in quantum states of an electron in a gated quantum dot in a single layer of TMDC as shown in Figure \ref{fig1}. This is necessary, since valleys and the spin-orbit coupling prevent us from identifying qubits with electron spin  states only.

Following Ref. \cite{Bieniek_Hawrylak_2020}, the Hamiltonian of an electron in a single layer of TMDC is a sum of the bulk Hamiltonian $H_b$ and quantum-dot confinement potential $V_{QD}$  \cite{Bieniek_Hawrylak_2018, Bieniek_Hawrylak_2020}. The potential $V_{QD}(\vec{r})$ is approximated here by a Gaussian   potential $V_{QD}(\vec{r})=-V_{0}\exp\left(-r^{2}/R_{QD}^2 \right)$, where $V_0$ is the potential depth and $R_{QD}$ is the quantum dot radius.  The electron quantum dot wavefunction $|\Phi^{s}\rangle$ for the electron state $s$ satisfies the Schr\"odinger equation \cite{Bieniek_Hawrylak_2018, Bieniek_Hawrylak_2020}:
\begin{equation}
 (H_{b}+V_{QD}(\vec{r})) \ket{\Phi^{s}}= E^s \ket{\Phi^{s}}.
\label{main:hamqdeq1}
\end{equation}
As explained in Ref. \onlinecite{Bieniek_Hawrylak_2020}, we define a large computational rhombus consisting of millions of metal atoms (sublattice A), and two layers of upper and lower chalcogen atoms (sublattice B). We retain only even metal orbitals and form an even combination of the upper and lower chalcogen $p$-orbital. We wrap the computational rhombus on a torus, apply the periodic boundary conditions and obtain a set of allowed k-vectors over which we diagonalize the bulk Hamiltonian $H_b$.  The sublattice A wavefunctions are expressed as a linear combination of even metal $d$-orbitals, with angular momentum two and $m_d=0,\pm 2$ and even combination of two, top and bottom, sulfur dimer $p$-orbitals with angular momentum one and $m_p=0,\pm 1$. The conduction band (CB) even wavefunction at each wavevector is a linear combination of simple even Bloch functions on the metal and sulfur sublattices $l$ ($l=1,..,6$)
\begin{equation}
\ket{\phi^{\textrm{CB,ev}}_{k\sigma}} = \sum_{l=1}^{6} A^{\textrm{CB,ev}}_{k\sigma, l} \ket{\phi^{\textrm{ev}}_{k,l}}\otimes\ket{\chi_{\sigma}},
\label{eq:4}
\end{equation}
where $\ket{\chi_{\sigma}}$ represents the spinor part of wavefunction and
\begin{equation}
\ket{\phi^{\textrm{ev}}_{k,l}}=\frac{1}{\sqrt{N_{\textrm{UC}}}} \sum_{\vec{R_{l}}=1}^{N_{\textrm{UC}}} e^{i\vec{k}\vec{R_{l}}} \varphi^{ev}_{l}\left(\vec{r}-\vec{R_{l}}\right)
\end{equation}
are simple Bloch functions built with  even orbitals  $\varphi^{ev}_{l}$ . $N_{UC}$ is the number of unit cells and $R_l$ define the position of even orbitals in the computational box. By diagonalising the 6 by 6 bulk Hamiltonian we obtain the bulk even energy bands $E_{k\sigma}^{\textrm{CB,ev}}$ and wavefunctions $ A^{\textrm{CB,ev}}_{k\sigma,l}$. 

\begin{figure}[ht]
\centering
\includegraphics[width=0.47\textwidth]{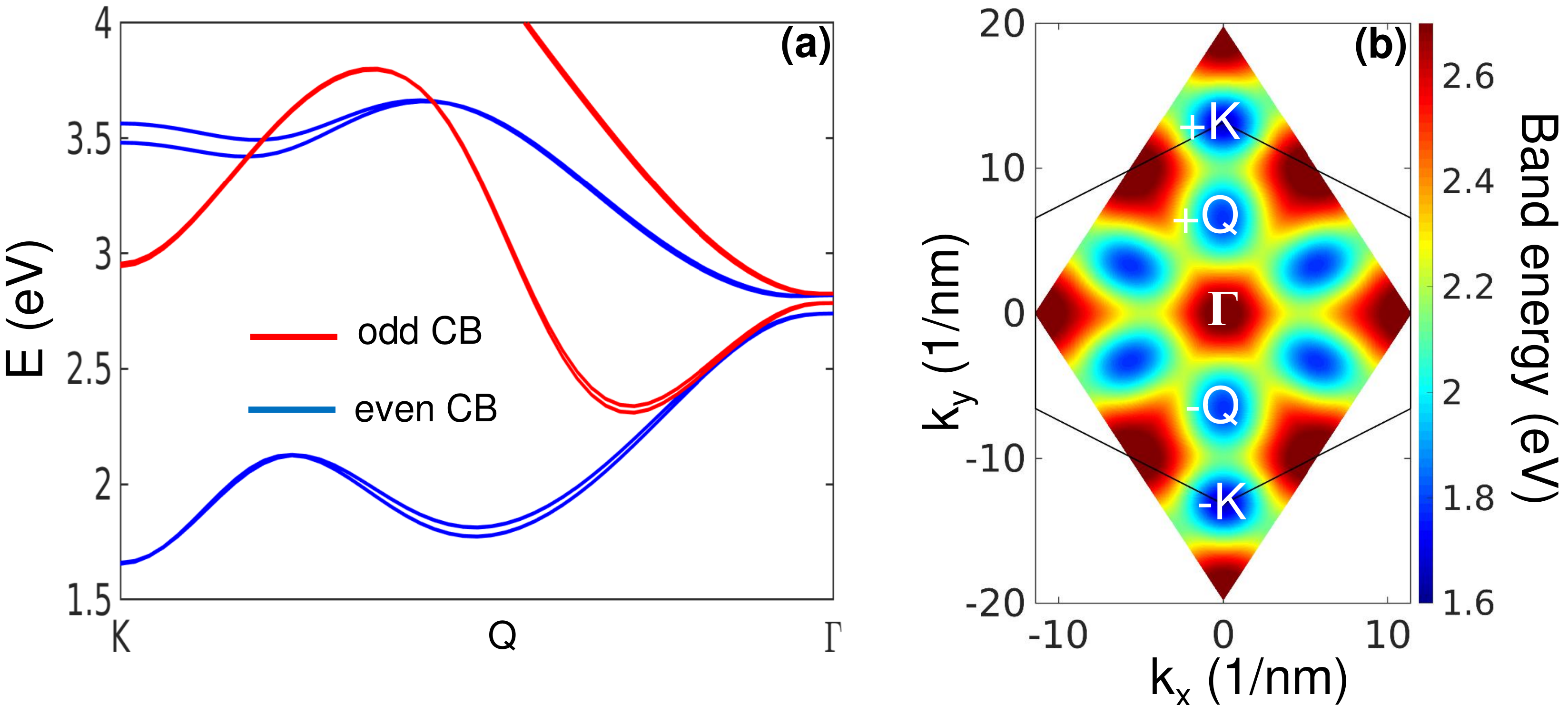}
\caption{(Color online) Bulk band structure of $\text{MoS}_\text{2}$. (a) Blue lines correspond to energy levels of even orbitals and red lines correspond to energy levels of  odd orbitals. (b) Lowest  conduction band energy levels on allowed values of  k points. $+K$ and $-K$ are global valley minima of the conduction band while the three $+Q$ and $-Q$ points correspond to 
local minima of the conduction band states.} 
\label{fig2}
\end{figure}
Figure \ref{fig2} shows the energy $E_{k\sigma}^{\textrm{CB}}$ of the lowest even conduction band (CB) as well as the map of the conduction band energies on the rhombus of $k$-space over which computations are carried out, including the $+K$ and $-K$ valley minima. The figure also contains the even conduction bands at a higher energy, to be discussed shortly.

In next step we expand the quantum dot wavefunction $\ket{\Phi^{s}}$ in terms of even lowest energy conduction band states given by Eq. \ref{eq:4}
\begin{equation}
\ket{\Phi^{s}}= \sum_{\vec{k}} \sum_{\sigma} B^{\textrm{s,CB,ev}}_{\vec{k}\sigma} \ket{\phi^\textrm{CB,ev}_{\vec{k}\sigma}}.
\end{equation}
The electron Schr\"odinger equation now converts to an integral equation for coefficients $B^{\textrm{s,CB,ev}}_{\vec{k}\sigma}$  
\begin{equation}
E_{q\sigma}^{\textrm{CB,ev}} B_{q\sigma}^{\textrm{s,CB,ev}} +\sum_{\vec{k}\sigma'} V_{q,k} A_{q\sigma,k\sigma'}B_{k\sigma'}^{\textrm{s,CB,ev}} = E^s  B_{q\sigma}^{\textrm{s,CB,ev}}.
\label{main:qdeq2}
\end{equation}
We see that the quantum dot confining potential in wavevector space turns out to be a product of  the lateral confinement $V_{q,k}$ and band contribution $A_{q\sigma,k\sigma'}$, with
\begin{equation}
\begin{aligned}
V_{q,k}=-V_0 \frac{S}{4\pi}R_{QD}^2 \exp\left(-\frac{(k-q)^2 }{4}R_{QD}^2\right)
\end{aligned}
\end{equation}
being the Fourier transform of the confining potential, with $R_{QD}$ being the radius of the quantum dot, $V_0$ $-$the depth of the gate potential, and $S$  $-$the reciprocal lattice unit-cell area. The band structure contribution to the scattering potential $A_{q\sigma,k\sigma'}$ is given by
\begin{equation}
A_{q\sigma,k\sigma'}=\sum_{l} \left(A^{\textrm{CB,ev}}_{q\sigma,l}\right)^{\dagger} \left(A^{\textrm{CB,ev}}_{k\sigma',l}\right).
\end{equation}

Solving the integral equation, Eq. \ref{main:qdeq2}, we obtain the quantum dot energy levels and wavefunctions. 
\begin{figure}[ht]
\includegraphics[scale=0.135]{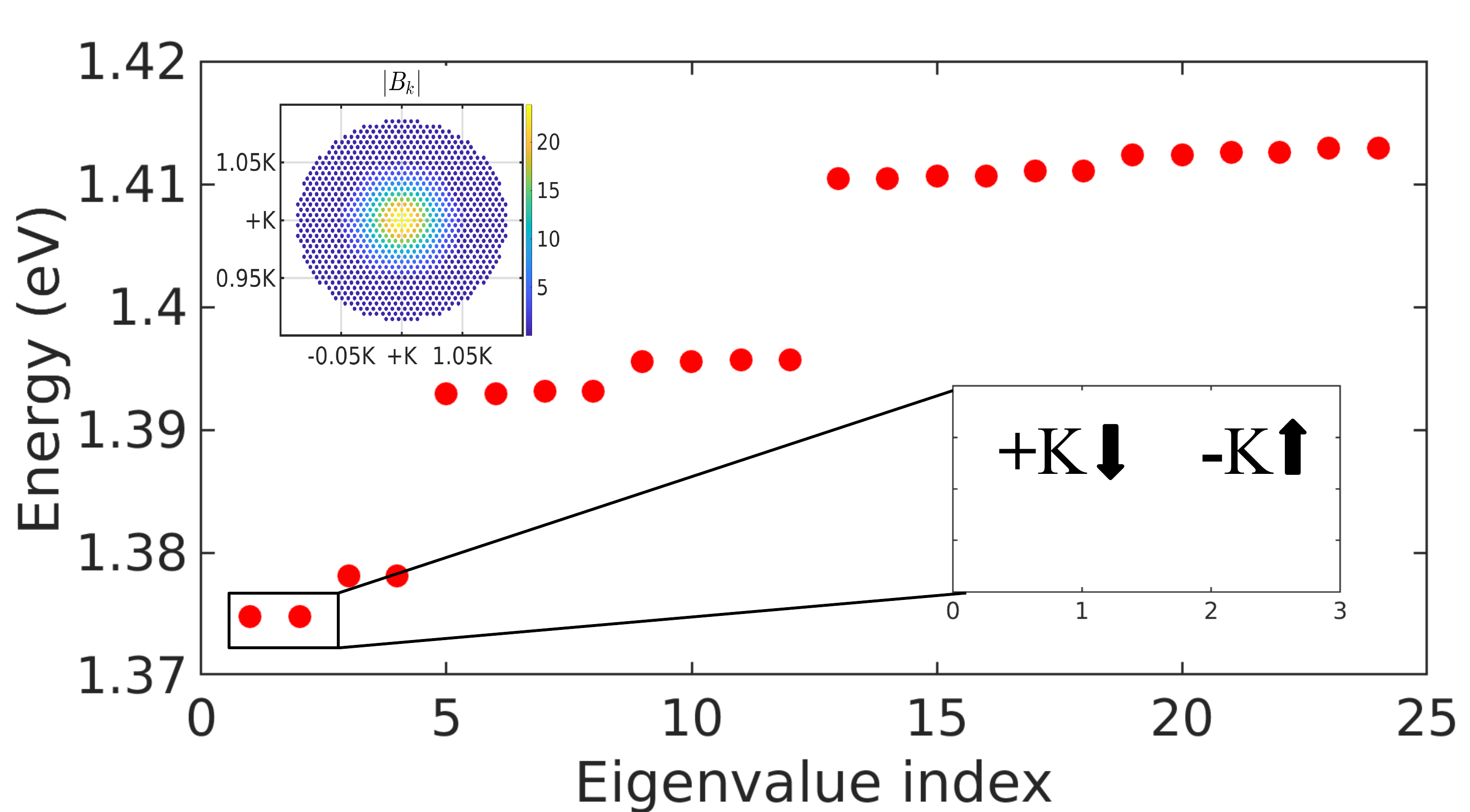}
\caption{(Color online) QD spectrum. Harmonic oscillator shell like  electronic states are formed due to the applied negative gate potential. Two  qubit states are indicated, well isolated from the rest of energy levels. Inset shows the highly localized qubit wavefunction in k-space. }
\label{fig3}
\end{figure}
Figure \ref{fig3} shows the  energy levels  of an electron confined in our quantum dot. We see that the levels are grouped into shells. 
The lowest energy shell consists of 4 low-energy states, related to 2 spin, up and down, states  and two valleys,  $+K$ and $-K$. The 4 states are split into pairs of levels by the spin-orbit interaction. The splitting, fully characterised in Ref. \cite{Bieniek_Hawrylak_2020}, is limited by the bulk value. From ab-initio calculations, the splitting is $~3$ meV for MoS$_2$ but splitting is greater than room temperature,  $30$ meV, for WSe$_2$\cite{Kadantsev_Hawrylak_2012,Bieniek_Hawrylak_2020}. We hence identify the two logical qubits, $\ket{0}$ and $\ket{1}$, with the two lowest energy levels, $\ket{0}=\ket{+K, \sigma=\downarrow}$  and $\ket{1}=\ket{-K, \sigma=\uparrow}$, shown in Figure 3. 
Additionally, the energy of the s states depends on the depth of the confining potential of the order of  300 meV used here. This guarantees stability of the qubit. The dependence of the energy separation of the qubit states from excited states on potential depth and radius were already addressed by our previous work\cite{Bieniek_Hawrylak_2020}. We want to note that, in our unpublished work we also studied the effect of impurity on the qubit states in the valence band . We found that the impurity shifted the energy of qubit states but preserved the valley-spin locking and degeneracy.

\section{Single qubit operations}
In this section we discuss the necessary steps for single logical qubit manipulation. 
Before rotating qubit states, the degeneracy of the valley-spin locked system should be lifted so that one can use the doublet state as a qubit, as can be seen from Figure 3. The degeneracy of spin-valley locked system can be lifted by applying a magnetic field ( for $\sigma_z$ operation), which is known as valley-Zeeman splitting \cite{wu2018valley, zajac2018resonantly}. This procedure will prepare and initialize the qubit states for $\sigma_x$ rotation. Hence, in order to rotate the qubit we need to be able to turn on both the $\sigma_z$ and $\sigma_x$ operations in the space of the logical qubit. 
To rotate the levels we need to be able to turn on the $\sigma_x$ operation. This operation needs to flip the logical qubit, i.e., induce a transition changing the spin and changing the valley. We will discuss this operation as composed of two steps, spin flipping and valley flipping.

\subsection{Vertical gate-valley conserving spin rotations}
Let us discuss how we can rotate the spin of an electron in a logical qubit  $\ket{0}=\ket{+K, \sigma=\downarrow}$ without changing the valley. We will accomplish this by turning on a vertical electric field $E_z$. The vertical electric field implies a higher potential $V_{E}/2$ on the upper sulfur layer and a lower potential $-V_{E}/2$ on the lower sulfur layer, with zero potential on the metallic layer. The applied bias acts primarily on sulfur layers and mixes the even combination of sulfur orbitals with odd combination of sulfur orbitals, and this mixes the even and odd conduction bands. Hence we need to determine the electronic states which are odd with respect to the metallic layer. There are two odd metal orbitals, $l=2, m_d=\pm 1$ on the sublattice A, and three odd sulfur dimer $m_p=0,\pm 1$ orbitals on the sublattice B. We expand the odd wavefunction in terms of odd metal and odd chalcogen dimer wavefunctions as:

\begin{equation}
\begin{aligned}
	\ket{\phi^{\textrm{CB,odd}}_{k\sigma}} = \sum_{l=1}^{5} A_{k\sigma,l}^{\textrm{CB,odd}} \ket{\phi_{k,l}^{\textrm{odd}}}\otimes \ket{\chi_{\sigma}}.
\end{aligned}
\end{equation}
The odd orbital Hamiltonian is obtained and diagonalised at each wavevector k. The lowest odd conduction band energy $E^{\textrm{CB,odd}}_{k}$ (red) is plotted in Figure \ref{fig1} together with the energy of even conduction bands (blue). We see that the odd band energy is higher than the even band by approximately $1$ eV. In order to understand all the steps we now retain only the lowest even and odd conduction band states and include both the spin-orbit coupling $V_{SO}$ and the odd-even orbital coupling $V_E$ by the applied electric field. 
In the presence of the electric field and the spin-orbit coupling, the bulk Hamiltonian can now be written in a block form as:
\begin{equation}
H = 
\begin{bmatrix}\label{main:bulkeq2}
H^{\textrm{ev}}_{\downarrow} & V_{\textrm{E}}^{\textrm{ev-odd}} & 0 & V_{\textrm{SO}\downarrow\uparrow} \\
  & H^{\textrm{odd}}_{\downarrow} & V_{\textrm{SO}\downarrow\uparrow} & 0 \\
  &   & H^{\textrm{ev}}_{\uparrow} & V_{\textrm{E}}^{\textrm{ev-odd}} \\
  &   &   & H^{\textrm{odd}}_{\uparrow}
\end{bmatrix}.
\end{equation}
We assumed here  that the applied electric field $E$ created  negative and positive voltages applied to lower and upper chalcogen atoms of bulk TMDC layer, respectively. The spin-orbit coupling in turn couples spin up and down states with even and odd metal orbitals, hence  the electric field and the spin-orbit coupling couple even and odd band states.

The vertical gate generates a laterally homogeneous electric field which couples odd and even orbitals of each chalcogen dimer. This translates into coupling of odd and even conduction bands at a given wavevector $k$. We assume the voltage $\hat{V}_{E}(z)$ due to applied electric field  such that  $-V_{E}/2$ is the voltage  applied on the chalcogen atom located on the lower layer, $ V_{E}(z=0)=0$ is the voltage on the metal layer, and $+V_{E}/2$ is the voltage on the chalcogen atom located on the upper layer of TMDC. The matrix element coupling the odd and even bands at each wavevector is given by:
\begin{equation}\label{matrixelement:eqB10}
\begin{aligned}
&V_{\textrm{E}}^{\textrm{ev-odd}}=\langle \phi^{\textrm{CB,odd}}_{k\sigma}|\hat{V}_{E}|\phi^{\textrm{CB,ev}}_{k\sigma} \rangle = \\
&\frac{1}{2N_{UC}}\sum_{\vec{R}_{B'},\vec{R}_{B}} \sum_{\substack{m_{p'},m_{p}\\=0,\pm1}} \left(A^{\textrm{CB,odd}}_{\vec{k}\sigma,m_{p'}}\right)^{*} \left(A^{\textrm{CB,ev}}_{\vec{k}\sigma,m_{p}}\right) e^{i\vec{k}\left( \vec{R}_{B} - \vec{R}_{B'} \right)}\\ 
& \times \iint dz d^{2}r \left[\varphi^{u}_{m_{p'}}(z, \vec{r}-\vec{R}_{B'}) -\varphi^{d}_{m_{p'}}(z, \vec{r}-\vec{R}_{B'})\right]^{*}\\
&\times \hat{V}_E(z)\times \left[ \varphi^{u}_{m_{p}}(z, \vec{r}-\vec{R}_{B}) + \varphi^{d}_{m_{p}}(z, \vec{r}-\vec{R}_{B}) \right].
\end{aligned} 
\end{equation}
We see that the only contribution to the matrix element comes from chalcogen orbitals on top (up) and bottom (down) chalcogen layers. The main contribution to this expression comes from combined  orbitals on upper and lower layers :
\begin{equation}\label{matrixelement:eqB10}
\begin{aligned}
&V_{\textrm{E}}^{\textrm{ev-odd}}= \frac{1}{{2 N_{UC}}} \times\\
&\sum_{\vec{R}_{B^{'}},\vec{R}_{B}}\sum_{\substack{m_{p'},m_{p}\\=0,\pm1}} \left(A^{\textrm{CB,odd}}_{\vec{k}\sigma,m_{p'}}\right)^{*}  \left(A^{\textrm{CB,ev}}_{\vec{k}\sigma,m_{p}}\right) e^{i\vec{k} \left( \vec{R}_{B}-\vec{R}_{B'}\right) }\\
&\times \bigg[ \iint  \,dz\,d^2r \varphi^{u *}_{m_{p'}}(z, \vec{r}-\vec{R_{B^{'}}})\left(\frac{V_E}{2}\right)\varphi^{u}_{m_{p}}(z, \vec{r}-\vec{R_{B^{}}}) \\
&- \iint  \,dz\,d^2r \varphi^{d *}_{m_{p'}}(z, \vec{r}-\vec{R_{B^{'}}})\left(-\frac{V_E}{2}\right)\varphi^{d}_{m_{p}}(z, \vec{r}-\vec{R_{B}}) \bigg].
\end{aligned}
\end{equation}
The integrals over $r$ and $z$ give $\delta{(R_B,R_{B}^{'})}$ and $\delta{(m_p,m_{p}^{'})}$. It is now clear that the final approximate result can be written simply as
\begin{equation}\label{matrixelement:eqVe}
\begin{aligned}
&V_{\textrm{E}}^{\textrm{ev-odd}} = \frac{V_E}{2} \sum_{m_{p}=0,\pm1} \left(A^{\textrm{CB,o}}_{\vec{k}\sigma,m_{p}}\right)^{*} \left(A^{\textrm{CB,ev}}_{\vec{k}\sigma,m_{p}}\right). 
\end{aligned} 
\end{equation}
We see that the electric field couples odd and even conduction bands and the magnitude of that coupling is proportional to a product of odd and even band amplitudes $A$ at each wavevector $k$, summed over all $m_p$ orbitals. However, the odd and even conduction bands have a different composition of chalcogen and metal orbitals at the bottom of the $+K$ valley. The selection rule derived in Ref. \onlinecite{Bieniek_Hawrylak_2018} implies that metal $m_d$ and chalcogen $m_p$ orbitals satisfy the selection rule $1+m_p-m_d=0,\pm 3$ . Hence the even band is built of $m_d=0$ and $m_p=-1$ orbital but the odd band is built on the $m_d=-1$ and $m_p=+1$ chalcogen orbitals. Chalcogen orbitals in the odd and even bands are different and the coupling strength, product of the same $m_p$ orbitals at the bottom of the $+K$ valley, vanishes. Hence the mixing of even and odd bands due to the vertical electric  field has a nontrivial dependence on the wavevector and so does the contribution to quantum dot states.

The mixing of odd and even bands for the same spin is only the first step in the spin rotation. Let us now turn our attention to the second step, induced by the spin-orbit coupling. The spin-orbit interaction is acting much more strongly on metal orbitals than calchogen orbitals. Starting with the even $m_d=0$ spin-down orbital, the spin-orbit interaction couples this state with the odd, $m_d=-1$ and spin-up orbital. Hence it is clear that the odd orbitals are needed to flip the spin. We can write spin orbit interaction mixing the lowest odd and even conduction bands on metal atoms (sublattice A) as: 
\begin{equation}\label{matrixelement:eqB11}
\begin{aligned}
& V_{SO\downarrow\uparrow} = \langle \phi^{\textrm{CB,ev}}_{k\downarrow} | \hat{V}_{SO} | \phi^{\textrm{CB,odd}}_{k\uparrow}\rangle = 
\\
&  \frac{1}{N_{UC}} \sum_{\vec{R}_{A},\vec{R}_{A'}} \sum_{\substack{m_{d}=0,\pm2\\m_{d'}=\pm1}}\left(A^{\textrm{CB,ev}}_{\vec{k}{\downarrow},m_{d} }\right)^{*}\left(A^{\textrm{CB,odd}}_{\vec{k}\uparrow,m_{d'}}\right) e^{i\vec{k}\left(R_{A}-R_{A'} \right)}\times\\
&\iint  \,dz\,d^2r \varphi^{\textrm{e}v*}_{m_{d}}(\vec{r}-\vec{R}_{A})  \langle\downarrow | \hat{V}_{SO}(z,r)|\uparrow \rangle  \varphi^{\textrm{odd}}_{m_{d'}}(\vec{r}-\vec{R_{A'}}).
\end{aligned} 
\end{equation}
The main contribution comes from the $ L\cdot S$ Thomas spin-orbit coupling on a given metal atom. Given that the CB is composed mainly of $m_d=0$ orbitals, results in expression:
\begin{equation}\label{matrixelement:eqB11b}
\begin{aligned}
V_{SO\downarrow\uparrow } &=\left(A^{\textrm{CB,ev}}_{\vec{k}\downarrow,m_{d=0}}\right)^{*}\left(A^{\textrm{CB,odd}}_{\vec{k}\uparrow,m_{d=-1}}\right) \times \\
&\langle\downarrow m_{d=0}|\hat{V}_{SO}|\uparrow m_{d=-1}\rangle.
\end{aligned}
\end{equation}
This SO term couples the even $m_d=0$ spin-down band with the $m_d=-1$ spin-up odd band, given by the product of amplitudes of the two bands weighted by the spin-orbit coupling matrix element.

Using the second-order perturbation theory in the basis of lowest even and odd conduction band states, we obtain the wavefunction of an electron in the valley $+K$ in the presence of both the electric field and the SO coupling. The wavefunction in the $+K$ valley with spin down $\Downarrow$ acquires a small admixture of the spin-up state:
\begin{equation}\label{main:bulkeq4}
\begin{aligned}
\Psi^{\textrm{CB,ev}}_{k,+K,\Downarrow} = \phi^{\textrm{CB,ev}}_{k\downarrow}\chi_{\downarrow} + D_{k\downarrow\uparrow}  \phi^{\textrm{CB,ev}}_{k\uparrow}\chi_{\uparrow}
\end{aligned}
\end{equation}
where, in the second-order perturbation theory 
\begin{equation}\label{main:bulkeq4}
\begin{aligned}
D_{k,\downarrow\uparrow}&=  \left( \frac{ V_{SO\downarrow\uparrow} V_E } {(\epsilon^{\textrm{CB,ev}}_{k,\downarrow} -\epsilon^{\textrm{CB,ev}}_{k,\uparrow}) (\epsilon^{\textrm{CB,ev}}_{k,\downarrow}-\epsilon^{\textrm{CB,odd}}_{k,\uparrow})} \right.
\\
&+ \left.  \frac{ V_E   V_{SO\downarrow\uparrow} }  
{(\epsilon^{\textrm{CB, ev}}_{k,\downarrow}-\epsilon^{\textrm{CB,ev}}_{k,\uparrow}) (\epsilon^{\textrm{CB,ev}}_{k,\downarrow}-\epsilon^{\textrm{CB,odd}}_{k,\downarrow})} \right).
\end{aligned}
\end{equation}
We see that the process of spin rotation is proportional to the applied vertical electric field and involves even and odd bands as well as the spin-orbit interaction. The same procedure can be applied to the $-K$ valley.

We can now return to quantum dot states and our logical qubit. The logical qubit state $\ket{0}=\ket{0,+K,\Downarrow}$ acquires a small spin-up component as
\begin{equation}
\begin{aligned}
\Phi^{0}_{+K,\Downarrow}
\cong \sum_{\vec{k} \in +K}  B_{\vec{k},\Downarrow}^{\textrm{CB,ev}} 
( \phi^{\textrm{CB,ev}}_{k,\downarrow} \chi_{\downarrow} + D_{k\downarrow\uparrow}
\phi^{\textrm{CB,ev}}_{k,\uparrow} \chi_{\uparrow})
\end{aligned} 
\end{equation}
while the logical qubit state $|1\rangle=|0,-K,\Uparrow \rangle$ acquires a small spin-down component
\begin{equation}
\begin{aligned}
\Phi^{1}_{-K,\Uparrow} 
 \cong \sum_{\vec{k} \in -K}  B_{\vec{k},\Uparrow}^{\textrm{CB,ev}} 
( \phi^{\textrm{CB,ev}}_{k,\uparrow} \chi_{\uparrow} + D_{k\uparrow\downarrow} \phi^{\textrm{CB,ev}}_{k,\downarrow} \chi_{\downarrow}).
\end{aligned} 
\end{equation}
We are now ready to couple the qubit states belonging to two different valleys.

\subsection{ Local gate - intervalley rotation}
We see that upon application of the vertical electric field the qubit states acquire admixtures of states in the same valley but with an opposite spin. We now introduce a local lateral gate  operator $\hat{G}$ which couples the two spin-valley-locked states  forming the qubit. The coupling defines the $\sigma_x$ matrix for logical qubit states:

\begin{equation}\label{matrixelement:eq2}
\begin{aligned}
&\langle 1|\sigma|0 \rangle=\langle\Phi_{-K,\Uparrow}| \hat{G} |\Phi_{+K,\Downarrow}\rangle
\\
&= \sum_{\vec{q}\in-K}  \sum_{\vec{k}\in+K}  B_{\vec{q},\Uparrow}^{{ev}^*} B_{\vec{k},\Downarrow}^{ev} G(q,k)  \times
\\
& ( D_{q\uparrow\downarrow}^*
 A^{ev}_{\vec{q},m_{d}=0\uparrow}  A^{ev}_{\vec{k},m_{d}=0\uparrow}   
 + D_{k\downarrow\uparrow}
 A^{ev}_{\vec{q},m_{d}=0\downarrow}  A^{ev}_{\vec{k},m_{d}=0\downarrow}).
\\
\end{aligned} 
\end{equation}
In what follows, we assume the local gate G to be a localized Gaussian given by its Fourier transform $G(q,k)=-G_0 \frac{R_{G}^2 { S}}{4\pi} \exp({\frac{-(\vec{k}-\vec{q})^2 R_{G}^2}{4}})$, where $R_{G} = 0.2$ nm is the width of Gaussian and $G_{0}= \, 1$ eV is its strength. Before proceeding to analyze the coupling matrix element, we can discuss the terms which significantly affect the strength of the coupling. One of the important terms is the QD wave function $B_{\vec{k}(\vec{q}),\Downarrow(\Uparrow)}^{ev}$ which is highly localized in $k$-space, as shown in the inset of Figure \ref{fig3}, where absolute value of the wavefunction of one of the qubit states is shown. As a result of the high localization in $k$-space, we can safely concentrate on states close to the bottom of the $+K(-K)$ valley.
Additionally, the energy differences between even-even and even-odd states are considered to be constant in this range and we take  $\epsilon^{\textrm{CB,ev}}_{k,\downarrow} -\epsilon^{\textrm{CB,ev}}_{k,\uparrow} \cong 3$ meV and $\epsilon^{\textrm{CB,ev}}_{k,\downarrow} -\epsilon^{\textrm{CB,odd}}_{k,\uparrow} \cong 1.3$ eV.

Also, the coupling of states is non-zero when the terms $D_{k(q)\downarrow\uparrow}$ are significant. These terms are proportional to the strength of the applied electric field and the spin-orbit coupling. While the spin-orbit coupling is the property of the material, the electric field can be turned on to activate the $\sigma_x$ matrix. The lateral gate $G$ is responsible for the coupling of the valley $+K$ and $-K$. The 
localized Gaussian potential $G(q,k)$ has to be a local perturbation with nonzero Fourier components $|\vec{q}-\vec{k}| \sim 2K$. When turned on, it will be responsible for flipping the valley index. We propose that a scanning tunneling microscope (STM) tip or a gated impurity could be used to realize this effect experimentally.  
\begin{figure}[ht]
\includegraphics[scale=0.16]{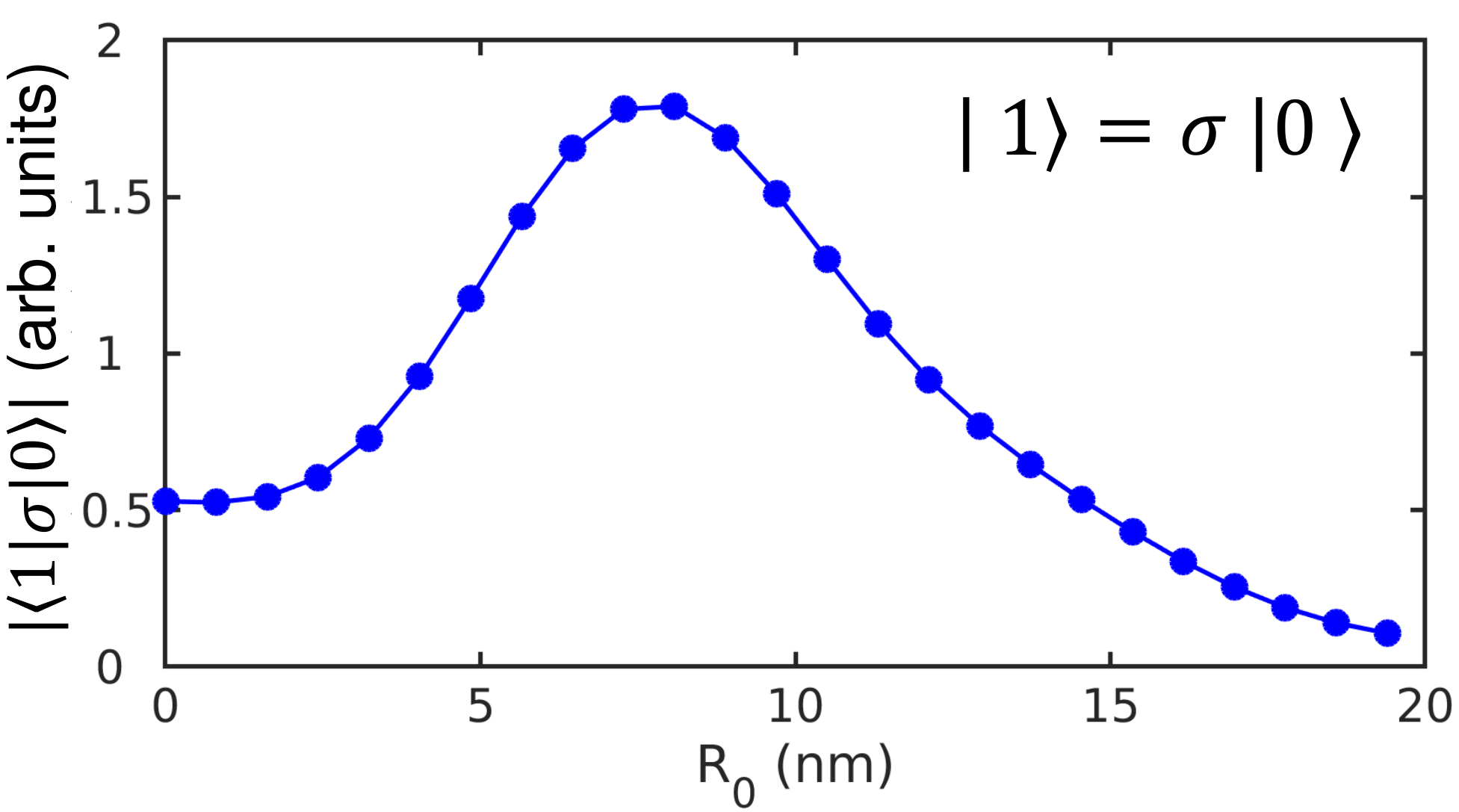}
\caption{(Color online) Logical qubit coupling matrix element as a function of position of the local gate $R_0$ for a given vertical electric field $V_E$.  }
\label{fig4}
\end{figure}
We  now discuss the behavior of the coupling matrix element as shown in Figure \ref{fig4}. A representative  QD studied in this project has a diameter of 40 nm centered at (x=0, y=0). We compute and plot the coupling matrix element as a function of the position  $\mathbf{R_{0}}$ of the local perturbation $G$ in a QD and $\mathbf{R_{0}}=R_{x} \hat{x}+ R_{y} \hat{y}$  where  $R_{x}$ (nm) $\in [0 ,20] $ and $R_{y}$(nm)$= 0 $ for a fixed applied vertical field $V_E$ (where we take $V_E$ = 1 eV in these calculations).
We move the perturbation $G$ from the center to the edge of the QD. The coupling matrix element has a finite value at the center of the QD 
and first increases  towards the halfway and decreases towards the edge of the QD. This nontrivial behavior can be traced to the nontrivial effect of the electric field on coupling of odd and even bands in TMDCs.

\section{Conclusions}
To summarise, we developed here a  theory of  valley-spin based qubits in gated quantum dots in a single layer of transition metal dichalcogenide. The qubits were identified with the two degenerate locked spin and valley states in a gated quantum dot.  The two qubit states were accurately described using a multi-million atom tight-binding model solved in k-space. The spin-valley locking and strong spin orbit coupling results in two degenerate states, one of the states of the qubit  being spin-down located at the +K valley, and the other state  located at the -K valley with spin up. We describe the gates necessary to rotate the spin-valley qubit as  a combination of the applied  vertical electric field enabling the spin orbit coupling in a single valley combined with a lateral strongly localised valley mixing gate. 
We note that suggested set up shown in Figure 1 can be readily implemented for one qubit operation. On the other hand, to be able to study manipulation of two or more qubit realizations, one can introduce impurity centers to mimic the role of the STM set up proposed in Figure 1. In addition, the aim of present work is to show how one can manipulate a single spin-valley qubit. The universal quantum computation requires also a two-qubit gate. Hence, our future work will focus on a microscopic description of a two-qubit gate operations where we will build on this and previous works \cite{david2018effective}.

\section*{Acknowledgments}
The authors thank L. Szulakowska, M. Cygorek, Y. Saleem,  J. Manalo , J. B. Chouinard, A. Bogan, A. Luican-Mayer and L. Gaudreau for valuable discussions. This research was supported by the NSERC Strategic grant QC2DM, NSERC Discovery grant and University of Ottawa Research Chair in Quantum Theory of Quantum Materials, Nanostructures and Devices. M.B acknowledges financial support from Polish National Agency for Academic Exchange (NAWA), Poland, grant PPI/APM/2019/1/00085/U/00001. Computing resources from Compute Canada are gratefully acknowledged.




%

\end{document}